\documentclass[prl,aps, superscriptaddress,twocolumn]{revtex4-1}
\usepackage{amsmath,amssymb}
\usepackage{graphicx}
\usepackage{wasysym}
\usepackage{amsfonts}
\usepackage{bm}
\usepackage{enumerate}
\usepackage{color}
\usepackage[resetlabels]{multibib}
\usepackage{epstopdf}
\usepackage{latexsym}
\usepackage[breaklinks,colorlinks = true,linkcolor = red,urlcolor=cyan,citecolor=red]{hyperref}
\usepackage[caption=false,singlelinecheck=false]{subfig}

\usepackage{times}
\newcommand{\bea}{\begin{eqnarray}}
\newcommand{\eea}{\end{eqnarray}}
\newcommand{\be}{\begin{eqnarray}}
\newcommand{\ee}{\end{eqnarray}}
\newcommand{\bw}{\begin{widetext}}
\newcommand{\ew}{\end{widetext}}

\newcommand{\la}{\langle}
\newcommand{\ra}{\rangle}

\begin{document}
\title{Discovery of new magnetic orders on pyrochlore spinels}
\author{GiBaik Sim}
\email{gbsim1992@kaist.ac.kr}
\affiliation{Department of Physics, Korea Advanced Institute of Science and Technology, Daejeon 305-701, Korea}
\author{SungBin Lee}
\email{sungbin@kaist.ac.kr}
\affiliation{Department of Physics, Korea Advanced Institute of Science and Technology, Daejeon 305-701, Korea}

\date{\today}
\begin{abstract}
Frustration in spin system can give rise to unique ordered states and as a consequence several physical phenomena are expected such as multiferroics, high temperature superconductors and anomalous hall effect. 
Here we report the ``new magnetic orders" induced by anisotropic spin exchanges on pyrochlore spinels as the interplay of spin orbit coupling and geometrical  frustration. Due to complicated superexchange paths of B-site spinels, we claim that anisotropic interaction between next-nearest neighbors play an important role. Based on the systematic studies of generic spin model, we argue that several classical spin states can be explored in spinel systems; 
local XY state, all-in all-out state, Palmer-Chalker state and coplanar spiral state. In addition, we reveal new types of magnetic phases with finite ordering wavevectors, labeled as {\it `octagonal (prism)'} state and {\it `(distorted) cubic'} states. When the {\it `octagonal prism'} state is stabilized, non-zero scalar spin chirality induces alternating net current in addition to finite orbital current and orbital magnetization even in Mott insulators. Finally, we also discuss the relevance of {\it  `distorted cubic'} state to the magnetic order of spinel compound GeCo$_2$O$_4$.  
\end{abstract}
\maketitle

Magnetic frustration originates from competing interactions between different spin exchanges. 
Despite the simple spin interactions such as Heisenberg or Ising type, lattice geometries can give rise to frustration, often termed as geometrical frustration.\cite{ramirez1994strongly,balents2010spin} 
Apart from geometrical frustration, anisotropic spin exchanges can also give rise to frustration due to competing interactions.\cite{dzyaloshinsky1958thermodynamic,moriya1960anisotropic,jackeli2009mott} In general, such anisotropic spin exchanges are expected for spin-orbit coupled system described by the total angular momentum $J$. 
When anisotropic spin exchanges meet geometrical frustration, exotic magnetism emerges in Mott insulators.

Magnetic properties of the B site spinel compounds (AB$_2$X$_4$) own such interplay forming a corner shared tetrahedra, pyrochlore lattice structure.\cite{bertaut1964structure,lee2010frustrated,takagi2011highly}  
Focusing on pyrochlore lattice structure, one can derive the most generic spin model on symmetry grounds and indeed such generic spin model for nearest-neighbor interactions has been studied. Especially for rare-earth pyrochlore magnets (A$_2$B$_2$O$_7$), A site rare-earth ions also form a pyrochlore lattice and their partially filled $4f$ electrons can give rise to non-negligible anisotropic spin exchanges between nearest-neighbors.\cite{gardner2010magnetic,thompson2017quasiparticle,ross2011quantum,lee2012generic} 
However, the situation in B-site spinels with 3$d$-5$d$ magnetic ions is quite different from the case of rare-earth pyrochlores. 
In spinels, the superexchange paths induce anisotropic exchanges between next-nearest-neighbor to play an important role, thus generic spin model for B-site spinel compounds is significantly distinct from ones studied before and one could expect new types of magnetic orderings.


In this paper, we study the frustrated spin model on a pyrochlore lattice and find new magnetic phases and their unique properties. Focusing on spin orbit coupled B-site spinels, we write down the generic spin exchange Hamiltonian derived from the Hubbard model with a pseudospin $j_{\text{\it eff}}\!\!=\!\!1/2$ Kramers doublet. Considering superexchange paths for neighbors between B site magnetic ions, we claim that next-nearest neighbor anisotropic exchanges generally dominate in determining the magnetic ground states. 
Within a classical spin approach, we employ the Luttinger Tisza method\cite{luttinger1946theory} and iterative minimization method (with system size upto $12 \times 12\times 12$) to investigate the magnetic phase diagram. 
Our key result is the discovery of {\it new magnetic orderings} described with finite ordering wavevectors $\boldsymbol{Q}$; `{\it octagonal (prism)}' state and `{\it (distorted) cubic}' states. We also argue their interesting consequences like orbital currents and relevance to spinel compounds GeCo$_2$O$_4$.  
Another main issue is that such generic spin model can also lead to many interesting magnetic phases with $\boldsymbol{Q}$=0 ordering; local XY, all-in all-out and Palmer-Chalker states.\cite{bramwell2001spin, gingras2014quantum, palmer2000order} We note that these $\boldsymbol{Q}$=0 phases originate from anisotropic spin exchanges between next-nearest neighbors, even in the absence of nearest-neighbor interactions or long range dipolar interactions. \\

{\bf Model} 
\label{sec:model}

When the B site of spinel is occupied by transition metal ions with partially filled 3$d$-5$d$ orbitals, cubic crystal symmetry splits $d$ orbitals into $t_{2g}$ and $e_{g}$ and spin orbit coupling further splits $t_{2g}$ orbitals into $j_{\text{\it eff}}\!=\!3/2$ quadruplet and $j_{\text{\it eff}}\!=\!1/2$ doublet described by isospin configuration. Within pseudospin $j_{\text{\it eff}}\!=\!1/2$ doublet, one writes down the simple Hubbard model. 
\bea
\mathcal H = \sum_{ij,\alpha \beta} c^\dagger_{i \alpha} \Big( t_{ij} \mathbb{I}+ i {\textbf{d}}_{ij} \cdot \boldsymbol{\sigma}  \Big)_{\alpha \beta} c_{j \beta} 
+U \sum_i n_{i\uparrow}n_{j\downarrow},
\label{eq:Hubbard}
\eea
where the first term represents electron hopping within pseudospin 1/2 manifold between sites $i$ and $j$, and the second term represents electron interaction.  
For B-site spinels with $j_{\text{\it eff}}\!=\!1/2$ doublet, the nearest neighbor (NN) interactions via X mediated superexchanges completely vanish. 
However, the next-nearest neighbor (NNN) interactions via X-A-X mediated superexchanges generate both finite spin dependent and independent terms ({\sl i.e.} $t_{ij}\!\neq\!0 $ and ${\textbf{d}}_{ij}\!\neq\!0 $ between the NNN sites $i$ and $j$.) Furthermore, it turns out that the magnitude of $ |{\textbf{d}}_{ij}| / t_{ij}$ is quite large, indicating the importance of spin dependent hopping between NNNs. (See Supplementary Information for details.)
In Mott insulating regime, therefore, 
we derive the following spin Hamiltonian with NNs and NNNs at large $U$.
\bea
{\mathcal H} = \sum_{\la ij \ra} J_{ij}^{' \alpha\beta}(\Omega) ~ S_i^\alpha S_j^\beta + \sum_{\la \la ij \ra \ra} J_{ij}^{\alpha\beta}(\phi,\theta) ~ S_i^\alpha S_j^\beta ,
\label{eq:general-spin-model}
\eea
where $\la ij \ra $ and $\la \la ij \ra \ra$ indicates the nearest-neighbor (NN) and next-nearest-neighbor (NNN) interactions between $i$ site and $j$ site, $J'$ and $J$ characterize exchange coupling between NNs and NNNs respectively. On symmetry grounds, the interactions between NNs only require one dimensionless parameters $\Omega\!=\!\tan^{-1} \Big( \pm | \textbf{d}_{ij} | /t_{ij} \Big) $ with $\la ij \ra$,   
while the  interactions between NNNs require two parameters $\phi$ and $\theta$; 
$\phi$ which parameterizes the unit vector $\hat{\textbf{d}}_{ij}$ lying on the plane perpendicular to the two fold rotation axis and $\theta\!=\!\tan^{-1} \Big( \pm | \textbf{d}_{ij} | /t_{ij} \Big) $ with $\la \la ij \ra\ra$. 
(Detailed derivation is explained in Supplementary Information.)
For clarification, Eq.\eqref{eq:Jmatrices} exemplifies $\textbf{J}_{ab}^{'}(\Omega)$ and $\textbf{J}_{cd}(\phi,\theta)$ for a given bond, marked as $J_{03}$ and $J'_{03}$ respectively in Fig.\ref{fig:pyrochlore}. (See Supplementary Information for detail form of ${j_n^{'}}$ and ${j_n}$ in terms of $\Omega$ and $(\phi, \theta)$ respectively.)
\bea
\textbf{J}_{03}^{'}(\Omega) \!=\! 
\left(\!
\begin{array}{ccc}
	j_2^{'} & j_3^{'} & j_4^{'} \\
	j_3^{'} & j_2^{'} & j_4^{'} \\
	\!-\!j_4^{'} & \!-\!j_4^{'} & j_1^{'} \\
\end{array}
\! \right)\!,~
\textbf{J}_{03}(\phi,\theta) \!=\! 
\left(\!
\begin{array}{ccc}
	j_6 & j_3 & \!-\!j_2 \\
	j_5 & j_6 & \!-\!j_4 \\
	j_4 & j_2 & j_1 \\
\end{array}
\! \right)
\label{eq:Jmatrices}
\eea

The lattice structure of pyrochlores especially for the connectivity between NNNs is shown in Fig.\ref{fig:pyrochlore}. The super tetrahedron includes four tetrahedra (colored in red) and each tetrahedron is made by four sublattices (green spheres). Here, the NNs are marked in orange solid lines. In the super tetrahedron structure, each face consists of a hexagon formed by three tetrahedra. In a hexagon (we focus on one at the bottom of the super tetrahedron.), the NNNs connect different sublattices forming two triangles, marked as blue solid lines in Fig.\ref{fig:pyrochlore}. In total, each sublattice has 6 NNs and 12 NNNs in a pyrochlore lattice. \\

\begin{figure}[t]
	\includegraphics[width=.7\columnwidth]{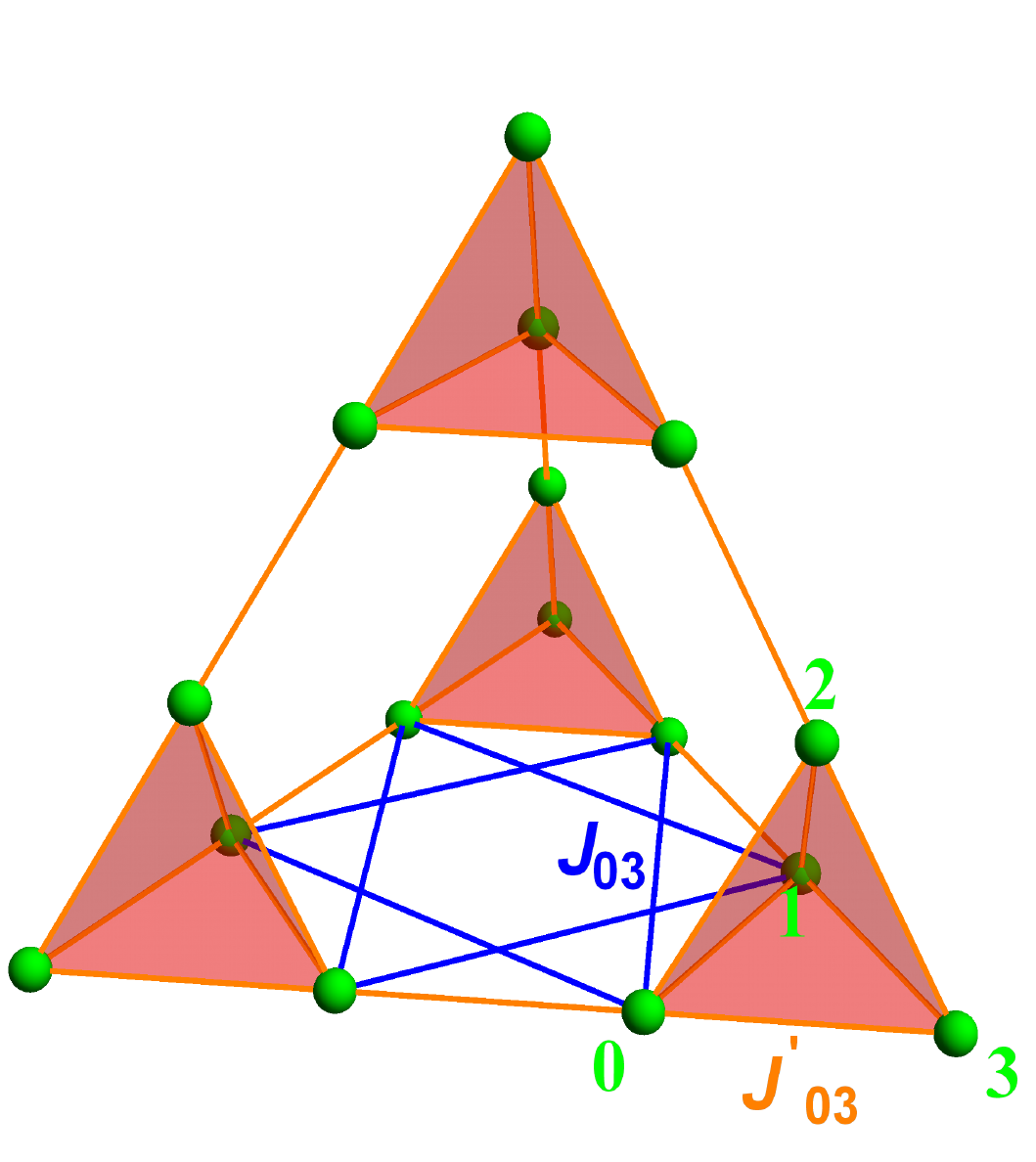}%
	\begin{picture}(45,0)
	\put(-50,100){\includegraphics[height=3.5cm]{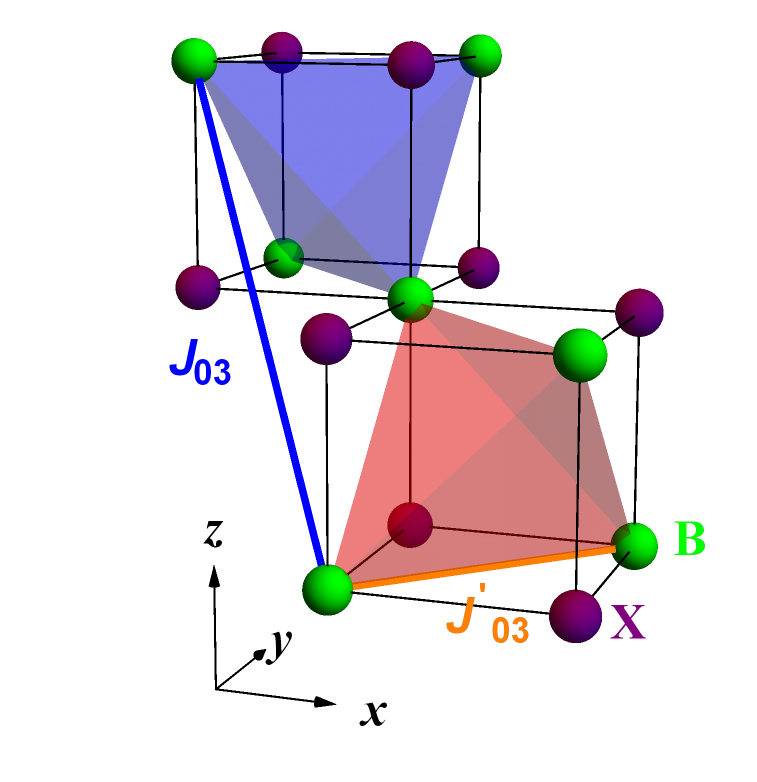}}
	\end{picture}
	\caption{(color online) Pyrochlore lattice structure and connectivity between nearest-neighbor sites (NN) and next-nearest-neighbor sites (NNN). To see the connectivity between NNNs, super tetrahedron is shown which consists of four tetrahedra. Each face of super tetrahedron form a hexagon connected by NNs (marked as solid lines with orange color). In every hexagon, two triangles (solid lines with blue color) exhibit the connectivity between NNNs. The inset shows schematic picture of B-site spinel structure only focusing on to X ions and B ions. See the main contents for detailed explanation.}
	\label{fig:pyrochlore}
\end{figure}

{\textbf{Magnetic Phase Diagram }}

\begin{figure}
	\includegraphics[width=0.9\columnwidth]{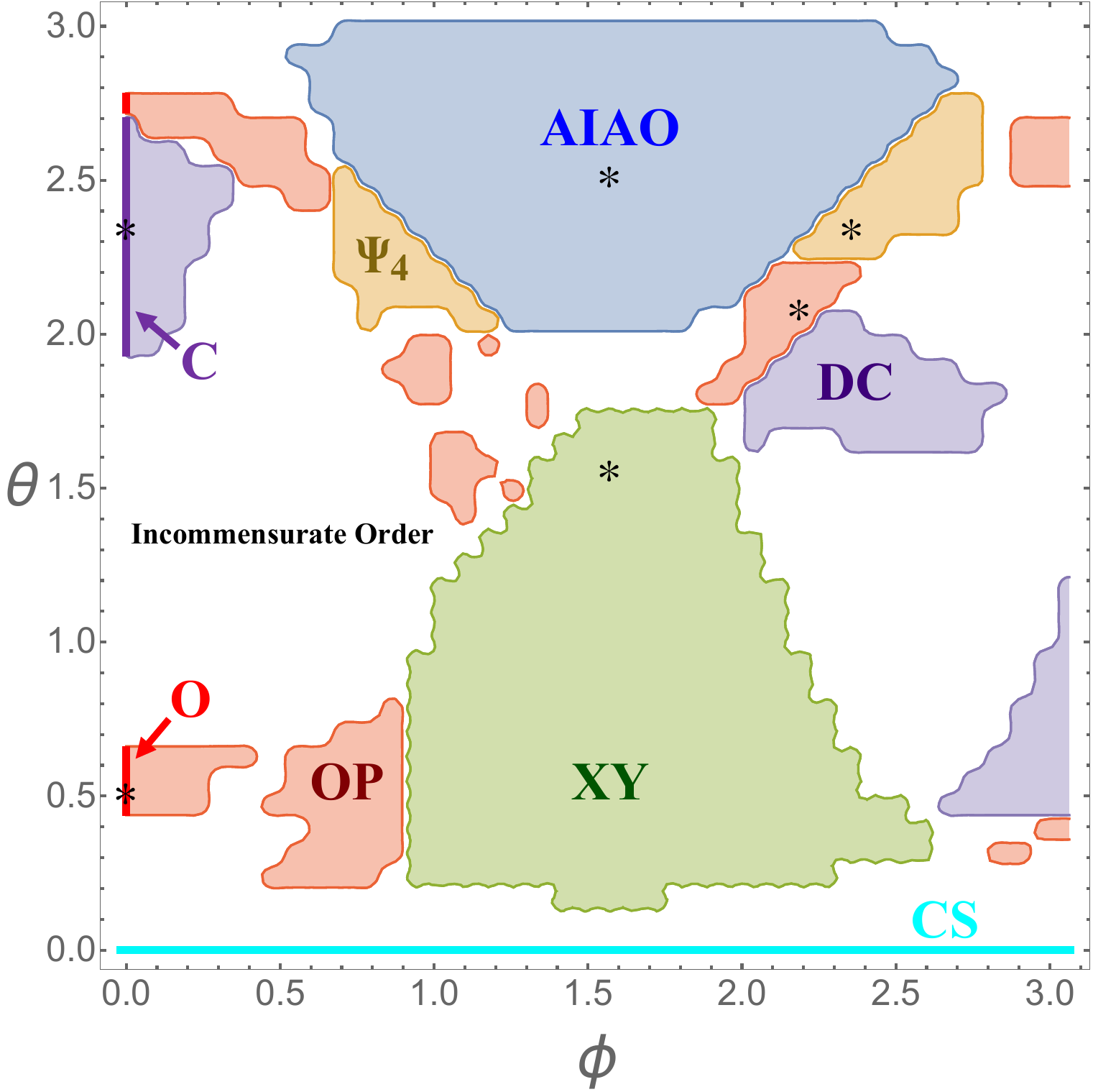}
	\caption{(color online) Phase diagram of magnetic ordering as functions of $\phi$ and $\theta$: Green,  blue and orange regions denote local XY phase ({\bf XY}), All-in-All-out phase ({\bf AIAO}) and Palmer-Chalker phase ($\mathbf{\Psi_4}$) respectively, which all belong to $Q\!=\!0$ ordering. Red and purple regions represent octagonal prism phase ({\bf OP}) and distorted cubic phase ({\bf DC}), which all belong to non-coplanar ordering with finite $Q$'s. Thick red, purple and cyan lines mark octagonal ({\bf O}), cubic ({\bf C}) and coplanar spiral ({\bf CS}) phases. See the main text for detailed description.
	}
	\label{fig:phase-diagram}
\end{figure}

Taking into account the previous argument of superexchange paths, we start from the generic spin model for NNNs {\sl i.e.} a finite $J_{ij}^{ \alpha \beta} (\phi,\theta)$ with $J_{ij}^{' \alpha\beta}(\Omega)\!=\!0$ first and then consider the stabilities of each phase with finite NN interactions. 
Fig.\ref{fig:phase-diagram} shows the phase diagram with two dimensionless parameters $\phi$ and $\theta$. 
It is worth to note that any point with a given parameter $(\phi,\theta)$ is identical to the point with $(\phi \!+\! m\pi,(-1)^m\theta \!+ \! n \pi)$ where $m$ and $n$ are arbitrary integers. Thus, we plot the phase diagram within the parameter ranges $\phi \! \in \! [0, \pi)$ and $\theta \! \in \! [0,\pi)$.

In the colored regions, we have found magnetic phases with commensurate ordering wavevectors where both Luttinger-Tisza and iterative minimization agrees with each other. (Explanation of each method is given in Supplementary Information.) However, in the rest of parameter space, we have only found incommensurate ordering wavevectors within Luttinger-Tisza approximation. To check the stability of each phase, one specific parameter at each phase is chosen  (marked as $*$ in Fig.\ref{fig:phase-diagram}) and we investigate the phase robustness in the presence of the nearest neighbor Heisenberg interaction $J_1$.   
Table.\ref{tab:summary} shows the summary of magnetic ground states we have found in our model.

\begin{table}
	\begin{tabular}{ |p{3cm}||p{1.7cm} | p{1.5cm} | p{1.3cm}|} \hline 
	Ground states & Coplanarity & $~~~~~~~ \boldsymbol{Q}$  & $ ~~~~ J_{1}(*)$ \\
	\hline \hline 
	Coplanar Spiral (CS) & Yes & $(\frac{4\pi}{3}00)$ & ~~~~~--- \\ \hline 
	Local XY (XY) & No & $(000)$ &$-1.8,\infty$ \\ \hline 
	All-in-All-out (AIAO) & No & $(000)$ &$-0.3,\infty$ \\ \hline 
	Palmer-Chalker ($\Psi_4$) & Yes & $(000)$ &  $-1.6,\infty$ \\ \hline 
	Octagonal (O) &Yes & $(2\pi\pi0)$ & $-2.6,0.2$\\ \hline 
	Octagonal Prism (OP) & No & $(2\pi\pi0)$ & $-0.1,0.9$ \\ \hline 
	Cubic (C) & No & $(\pi\pi\pi)$ &~~~~~---  \\ \hline 
	Distorted Cubic (DC) & No & $(\pi\pi\pi)$ &  $-0.7,10$  \\ \hline 
\end{tabular}
	\caption{ Summary of classical magnetic ground states of Eq.\eqref{eq:general-spin-model} : For given ground states, coplanarity and ordering wavevector, $\boldsymbol{Q}$, are described and the stability of each phase (at $*$ points in Fig.\ref{fig:phase-diagram}) in the presence of nearest-neighbor Heisenberg interaction $J_1$ is shown in the last column. (negative $J_1$ is for ferromagnet)}
\label{tab:summary}
\end{table}

{\bf \textit{Coplanar Phases with $\boldsymbol{Q}$$\neq$0 ---}}
At $\theta\!=\!0$, our model becomes a simple antiferromagnetic (AF) Heisenberg model of NNNs. In this limit, the model has been already investigated and its magnetic ground state is described by coplanar spiral phase (marked as {\bf CS} in Fig.\ref{fig:phase-diagram}.) with an ordering wavevector $\boldsymbol{Q}$$~\!\!=\!(\frac{4\pi}{3},0,0)$\cite{lapa2012ground}. Unlike the case of AF interaction between NNs where the system possess extensive degeneracy of magnetic ground states, AF interaction between NNNs favors specific coplanar spiral state with a finite $\boldsymbol{Q}$.  

\begin{figure*}
\begin{widetext}
%
		\begin{center}
			\subfloat[{\bf COP} of {\bf O}]{\label{fig:octagonal}\includegraphics[scale=0.22]{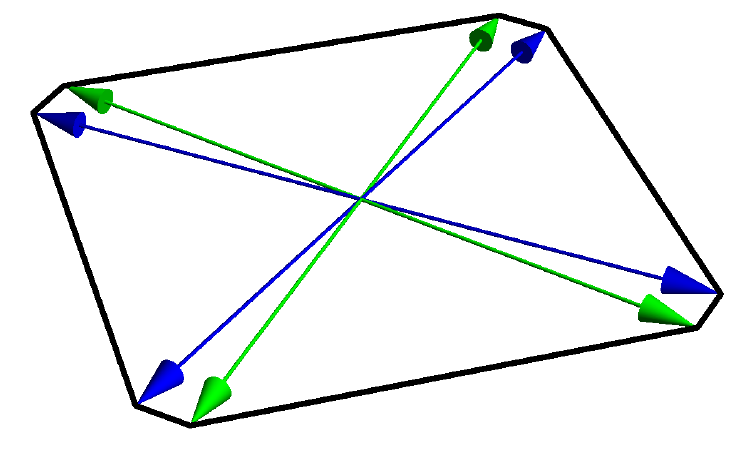}}~~~
			\subfloat[{\bf COP} of {\bf OP}]{\label{fig:octagonal-prism}\includegraphics[scale=0.22]{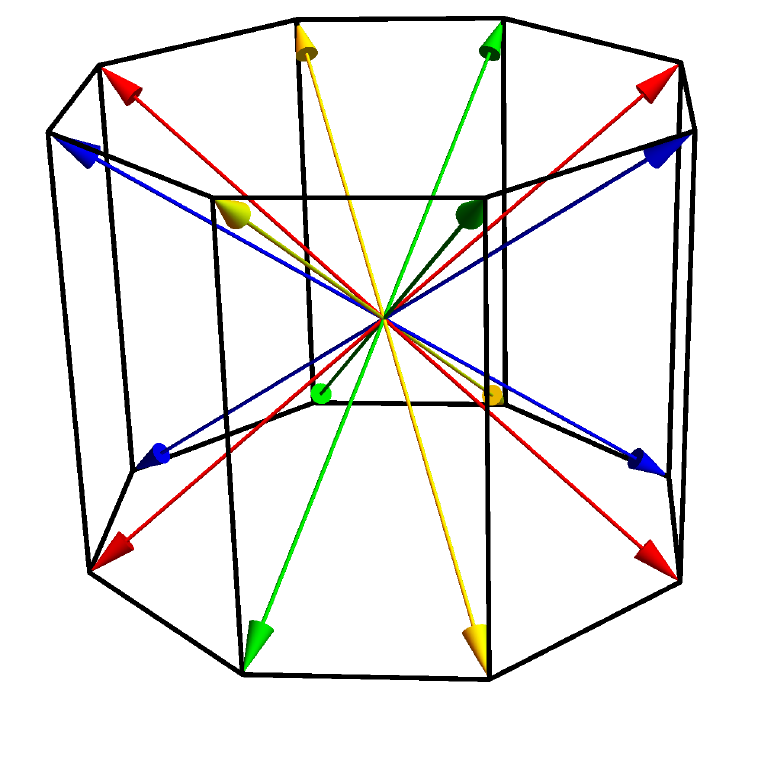}}~~~
			\subfloat[${\boldsymbol Q}$ of {\bf O} or {\bf OP}]{\label{fig:o-BZplot}\includegraphics[scale=0.15]{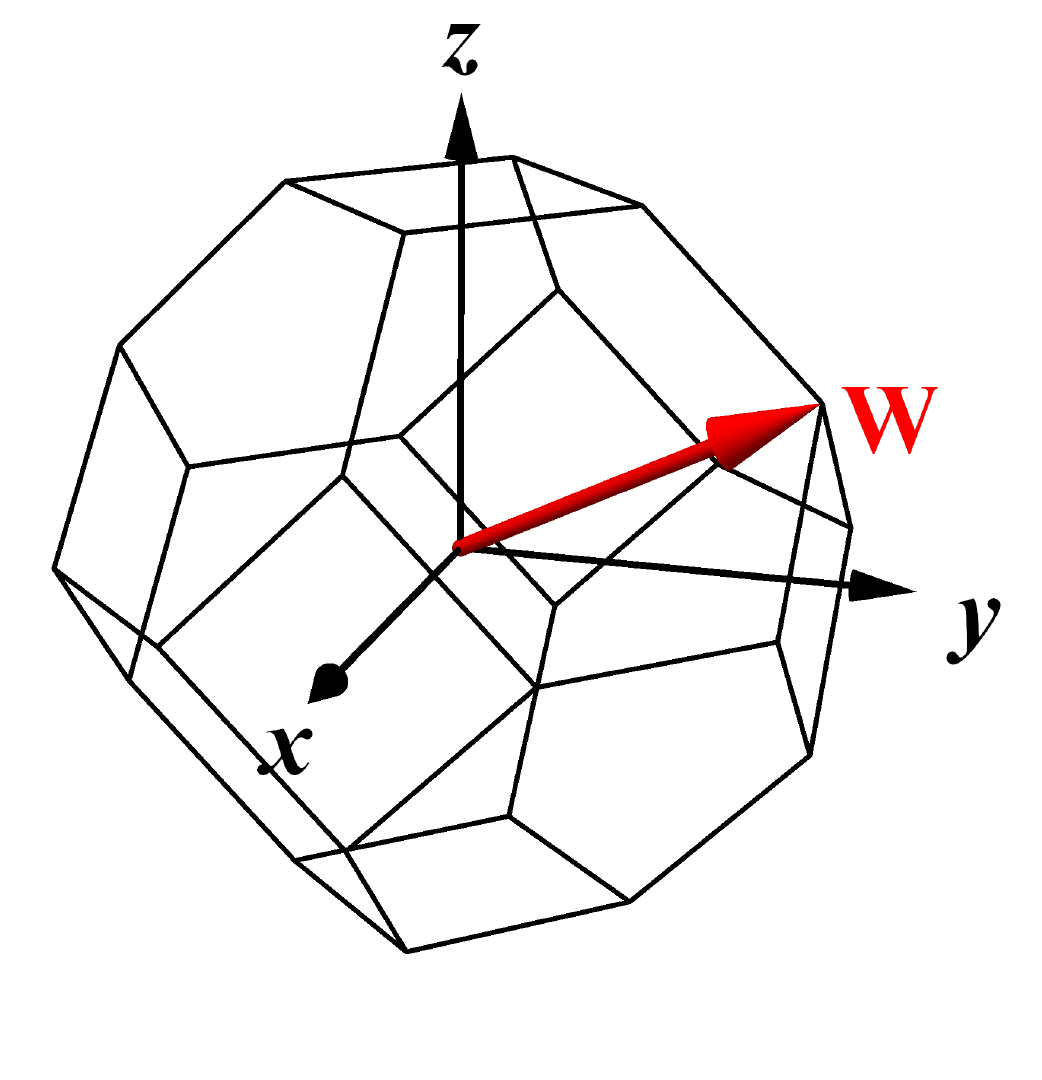}}
			\subfloat[{\bf COP} of {\bf C}]{\label{fig:cubic}\includegraphics[scale=0.22]{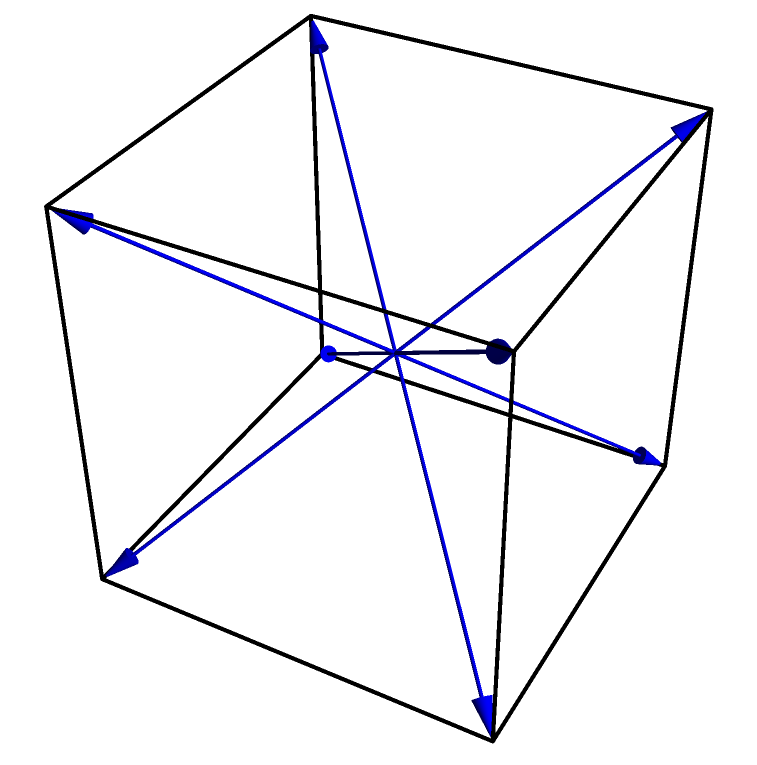}}~~~
                          \subfloat[{\bf COP} of {\bf DC}]{\label{fig:distorted-cubic}\includegraphics[scale=0.22]{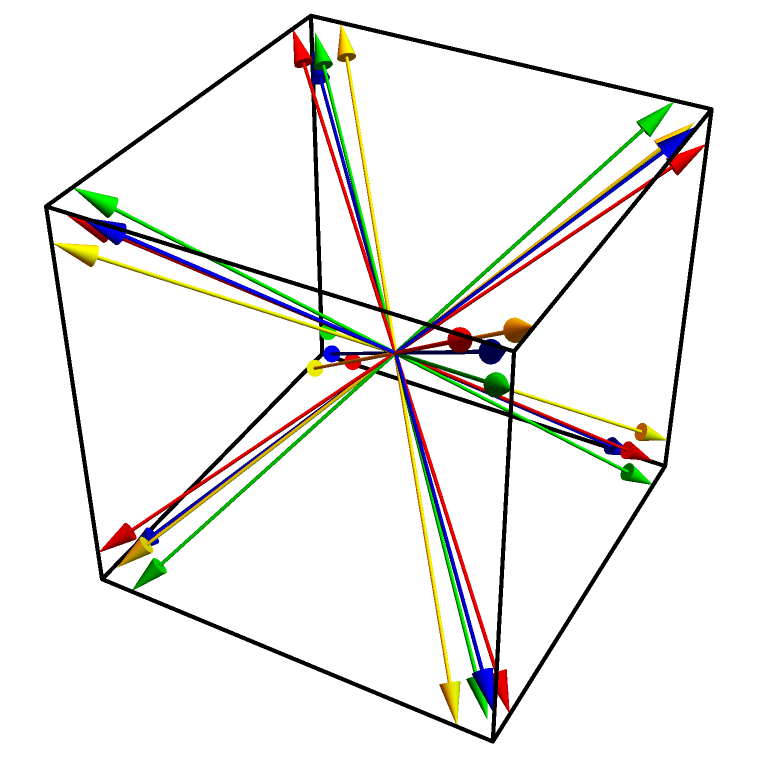}}
                          \subfloat[${\boldsymbol Q}$ of {\bf C} or {\bf DC}]{\label{fig:c-BZplot}\includegraphics[scale=0.15]{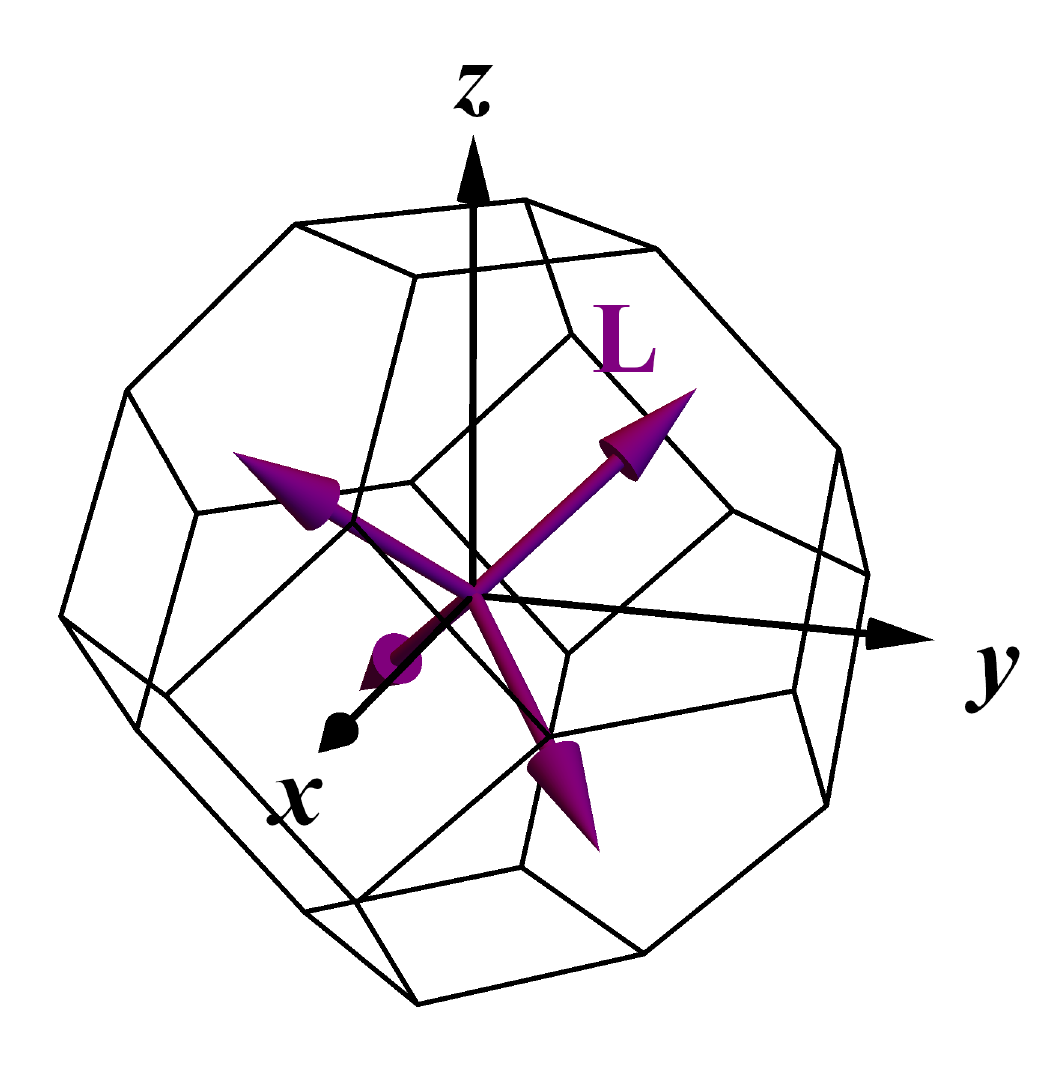}}
\caption{(Color Online) Common origin plots ({\bf COP}) and magnetic ordering wavevectors ${\boldsymbol Q}$ of {\it `octagonal (prism)'} phase and {\it `(distorted) cubic'} phase : 
(a) {\it `octagonal'} ({\bf O}) phase -- Spins point 8 different directions on a plane. Spin directions at sublattices 0 and 2 (1 and 3) are identical as colored in blue (green).
(b) {\it `octagonal prism'} ({\bf OP}) phase -- Spins at each sublattice form a plane and the planes are distinct for sublattices, forming a shape of an octagonal prism. Four distinct colors indicate the spin directions in different sublattices. (c) ${\boldsymbol Q}$ of {\bf O} and {\bf OP} phase -- Magnetic ordering wavevector is located at $W$ points shown in the 1st Brillouin zone. 
(d) {\it `cubic' phase} ({\bf C}) -- Spins point all 8 different directions forming a cubic structure. At every sublattice, spins are pointing every 8 vertices of cube. 
(e) {\it `distorted cubic' phase} ({\bf DC}) -- Spins point all 32 different directions resulting in distortion of perfect cube. At every sublattice, spins are pointing every 8 vertices of one rhombohedron.  
(f) ${\boldsymbol Q}$ of {\bf C} and {\bf DC} phase -- Both {\bf C} and {\bf DC} phases are described by ordering wavevectors located at high symmetric $L$ points. More detailed explanation is given in the main text.}
\end{center}
\end{widetext}
\end{figure*}

When $(\phi \!= \! 0,0.45\lesssim \! \theta \! \lesssim0.65)$ or $(\phi \!= \! 0,2.73 \! \lesssim \! \theta\lesssim2.77)$ marked as thick red colored region in Fig.\ref{fig:phase-diagram}, there is another coplanar phase which is described with an ordering wavevector at high symmetric $W$ point (See Fig.\ref{fig:o-BZplot}.). 
In this phase, spins at sublattice 0 and 2 (colored blue in Fig.\ref{fig:octagonal}) point 4 different directions, whereas at sublattice 1 and 3 (colored green in Fig.\ref{fig:octagonal}), spins point another 4 different directions. Since spins point all 8 different directions in total, we label this phase as {\it `octagonal'}({\bf O}) phase. The spin configuration at a sublattice $a$ is $\boldsymbol{S}_a \!=\! \boldsymbol{\Psi}_a \exp(i\boldsymbol{Q} \cdot\boldsymbol{R})+ c.c$ with $\boldsymbol{\Psi}_a\!=\!(f e^{-i\beta (-1)^a}\!,0,\!-i f e^{i\beta(-1)^a})$, $(f,\beta)$ are functions of the parameter $\theta$ and $\boldsymbol{Q}\!=\!(2\pi,\pi,0)$.

{\bf \textit {Magnetic Phases with $\boldsymbol{Q}$=0 ---}}
As we tune parameters $\phi$ and $\theta$, our model shows three different types of magnetic phases without enlarging unit cell, $\boldsymbol{Q}$$~\!\!=\!0$;
{\it All-in-All-out phase} ({\bf AIAO}), {\it Local XY phase} ({\bf XY}) and {\it Palmer-Chalker phase} ({$\mathbf{\Psi_4}$}) . All these phases belong to the degenerate ground state manifolds of AF Heisenberg interaction between NNs, satisfying the sum of four spins for every tetrahedron to be zero. Thus, these phases are stable even in the limit of $J_1 \! \rightarrow \! \infty$ as shown in Table.\ref{tab:summary}. 

In blue colored region in Fig.\ref{fig:phase-diagram}, {\it all-in-all-out} phase is stabilized. With a given local coordinate at each sublattice, all-in-all-out phase is described by all four spins in a unit cell aligned on their local $\hat{z}$ or$-\hat{z}$ axes, having two fold degeneracy. (See Supplementary Information for description of local coordinates.) 


In {\it local XY phase}, on the other hand, each spin lies in its local $xy$ plane, having the same azimuthal angle $\varphi_a \equiv \varphi$ in each sublattice. 
In this phase, one can freely choose $\varphi \in[0,2\pi)$ thus U(1) symmetry is present at a mean field level. (See the green colored region in Fig.\ref{fig:phase-diagram}.)
There is possible quantum or thermal order by disorder in this phase, having {\it pseudo}-Goldstone mode with little gap at low temperature. Since this local XY phase is stabilized by generic spin interactions between NNNs, it may choose different ordered phase due to fluctuations compared to the case studied in Refs.\onlinecite{savary2012order,wong2013ground,mcclarty2014order} and further detailed studies are required in the future. 

The Palmer-Chalker state (or equivalently referred as $\Psi_4$ state) is stabilized in the orange colored region in Fig.\ref{fig:phase-diagram}.\cite{palmer2000order,poole2007magnetic}
It is described by the following three different sets of azimuthal angles $\varphi_a$ for sublattices $a$ ($a \in$ 0,1,2 and 3), and their time reversal symmetric partners. (6 different configurations in total.); $\Big(\varphi_0\!\!=\!\!\varphi_1\!\!=\!\!\frac{\pi}{2}$, $\varphi_2\!\!=\!\!\varphi_3\!\!=\!\!\frac{3\pi}{2}\Big)$, $\Big(\varphi_0\!\!=\!\!\varphi_2\!\!=\!\!\frac{7\pi}{6}$, $\varphi_1\!\!=\!\!\varphi_3\!\!=\!\!\frac{\pi}{6}\Big)$,$\Big(\varphi_0\!\!=\!\!\varphi_3\!\!=\!\!\frac{11\pi}{6}$, $\varphi_1\!\!=\!\!\varphi_2\!\!=\!\!\frac{5\pi}{6}\Big)$ \cite{mcclarty2014order}. 
It is remarkable that this phase is stabilized by anisotropic NNN exchanges even without long range dipolar interaction or thermal/ quantum fluctuations.

\textit{ \textbf{Non-Coplanar Phases with Q$\neq$0 --- }}
Many non-coplanar magnetic orderings give fertile grounds where interesting phenomena emerge, such as orbital currents and anomalous Hall effect induced by non-zero spin chirality.\cite{taguchi2001spin,bulaevskii2008electronic} 
Although there exist earlier studies related to non-coplanar phases with $\boldsymbol{Q}\neq0$ in pyrochlores, those are stabilized only as metastable phases or due to itinerant electrons.\cite{chern2010noncoplanar,okubo2011cubic} In our model, we find two new types of magnetic orderings as ground states; {\it `octagonal prism'} phase and {\it `(distorted) cubic'} phase. In Fig.\ref{fig:phase-diagram}, both red or purple regions are where such non-coplanar states are stabilized.

{{\it Octagonal Prism Phase} ({\bf OP}) ---}
This non coplanar phase is described by the ordering wavevector $W$ same as coplanar {\it `octagonal'} phase but spins point 16 different directions forming a shape of an octagonal prism as shown in Fig.\ref{fig:octagonal-prism}. Thus we refer it as {\it `octagonal prism'} phase. In this phase, 
the spins at each sublattice still form a coplanar state but their planes are distinct by sublattices. Common origin plot of all spin configurations is shown in Fig.\ref{fig:octagonal-prism}. Red, green, yellow and blue colors indicate the spin directions in each sublattice. 
The magnetic unit cell is quadrupled where spins at each sublattice align on a certain plane but the planes are distinct for different sublattices. In details, spin configuration at a sublattice $a$ is parametrized as
$\boldsymbol{S}_a \!=\! \boldsymbol{\Psi}_a \exp(i\boldsymbol{Q} \cdot\boldsymbol{R})+ c.c$ with $\boldsymbol{\Psi}_a\!=\!(fe^{-i\beta(-1)^a}\!,\!ge^{-\frac{i \pi}{4}{(2a-1)}}\!,\!-i fe^{i\beta(-1)^a})$, $(f,g,\beta)$ are functions of $(\phi,\theta)$ and 
$\boldsymbol{Q}\!=\!(2\pi,\pi,0)$. 

\begin{figure}
	\includegraphics[width=0.9\columnwidth]{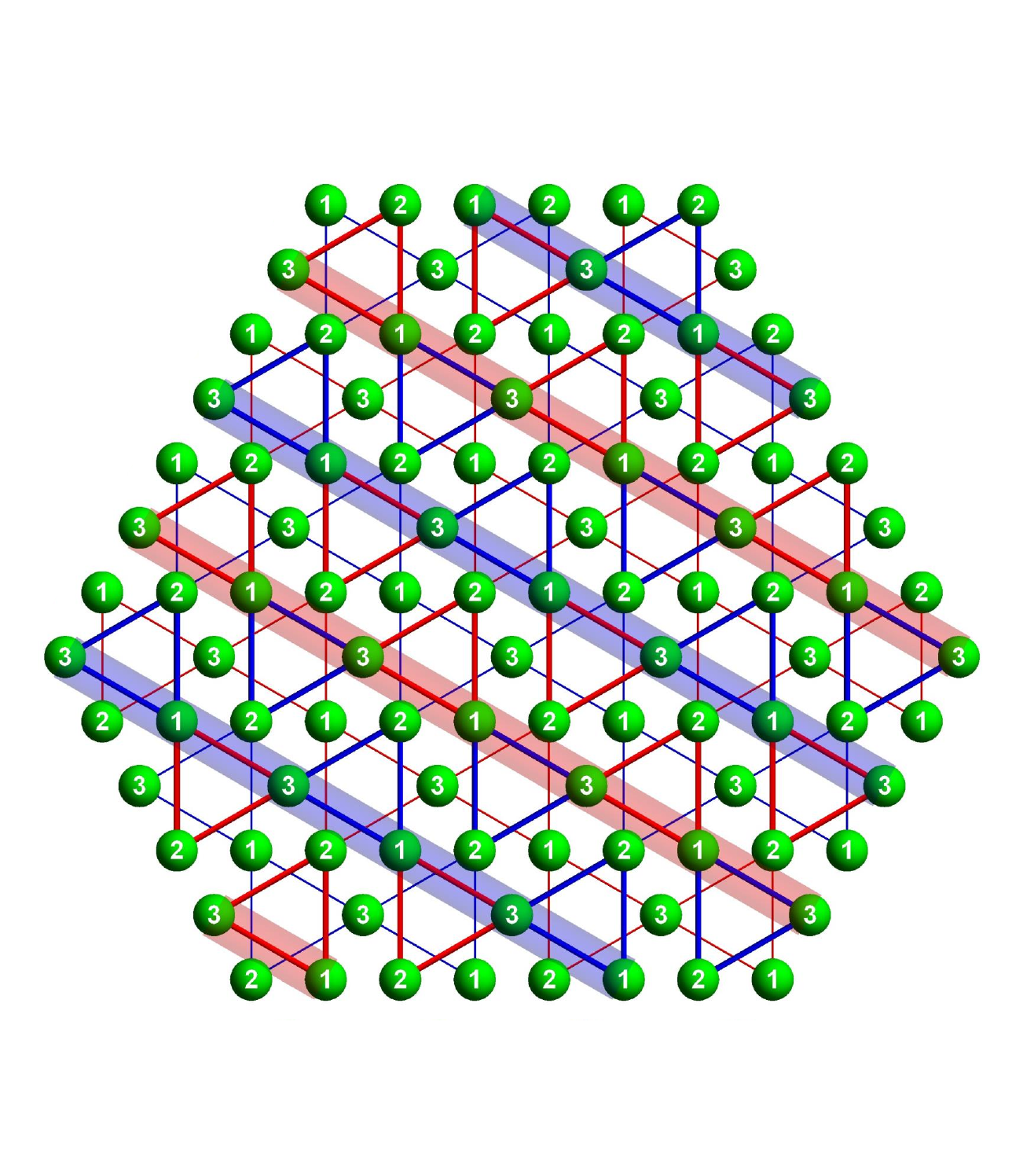}%
	\begin{picture}(200,0)
	\put(-40,200){\includegraphics[height=1.3cm]{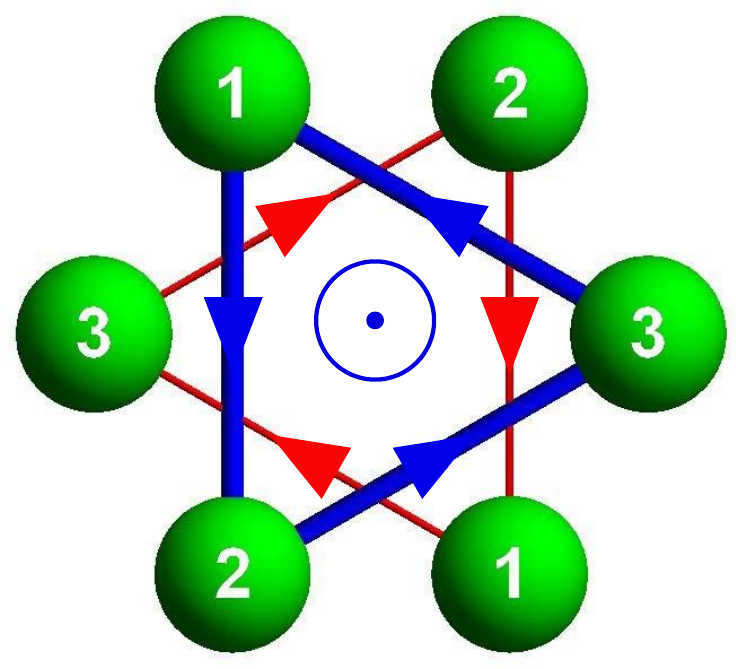}}
	\end{picture}
	\caption{(Color online) Scalar spin chirality at each triangle on (111) plane of pyrochlores. Blue (Red) colored triangles exhibit positive (negative) chirality and bond thickness represents its magnitude. The thick lines colored by blue and red are the direction $\pm [1\bar{2}1]$ of current channel induced by scalar spin chirality. See the main text for more explanation. 
	}
	\label{fig:orbital-current}
\end{figure}

In this phase, non-coplanar spin orderings can give rise to net scalar spin chirality on each triangle connected by the NNNs as shown in Fig.\ref{fig:orbital-current}. Having a finite spin chirality, the orbital current is generally induced even in the Mott insulating phase.\cite{shindou2001orbital,bulaevskii2008electronic} Fig.\ref{fig:orbital-current} shows the existence of net scalar spin chirality on each triangle formed by three sites $i$, $j$ and $k$, $\chi_{ijk} = {\bf S}_i \cdot ( {\bf S}_j \times {\bf S}_k )$ on (111) plane. Triangles with blue (red) color exhibit positive (negative) spin chirality $\chi_{ijk}$ where sites $i$, $j$ and $k$ are taken in counterclockwise direction, and the relative magnitudes of $\chi_{ijk}$ are represented by  bond thickness. At a given bond connected by two sites $i$ and $j$, the third order of perturbation theory at $\mathcal{O} \Big( t^3/U^2 \Big)$ induces the net current proportional to ${\bf I}_{ij,k} \!\sim\! \hat{\bf r}_{ij} ~ \chi_{ijk}$, where $\hat{\bf r}_{ij}$ is the unit vector from site $i$ to site $j$.\cite{bulaevskii2008electronic} 
This orbital current is like the persistent current in superconducting phase and it furthermore leads to the finite orbital magnetic moment ${\bf L}_{ijk} \propto \chi_{ijk} \hat{\bf z}$ where $\hat{\bf z}$ is normal to the plane of the triangle. As seen in Fig.\ref{fig:orbital-current}, the orbital current induced by net scalar spin chirality also makes the special current channel along $\pm [1\bar{2}1]$ directions (blue or red colored thick lines) on $(111)$ plane. Experimental observation of such spatially varying current channel will be the challenging future work. 

{{\it (distorted) Cubic Phase} ({\bf C} or {\bf DC}) ---}
At the region marked with thick purple color in Fig.\ref{fig:phase-diagram}, the ordering wavevectors are at high symmetric $L$ points (See Fig.\ref{fig:c-BZplot}.) and common origin plot of spin orderings is shown in Fig.\ref{fig:cubic}. In this phase, spins at every sublattice point 8 different directions forming a cubic structure and spin orderings at different sublattices are all identical consisting of the same cube. Thus we label it as {\it `cubic'}({\bf C}) phase. 
In details, spin configuration at each sublattice is represented as following.
\bea
\boldsymbol{S}_a= \frac{e^{ i \gamma_a}}{\sqrt{3}} \Big( \hat{\boldsymbol{x}}~ e^{i \boldsymbol{Q}_{a}\cdot\boldsymbol{R}  } 
+ \hat{\boldsymbol{y}} ~ e^{i\boldsymbol{Q'}_{a} \cdot\boldsymbol{R}}+
+ \hat{\boldsymbol{z}} ~ e^{i\boldsymbol{Q''}_{a}\cdot\boldsymbol{R}} \Big).
\label{eq:cubic}
\eea
Here, $\Big( \boldsymbol{Q}_a, \boldsymbol{Q'}_a, \boldsymbol{Q''}_a  ,\gamma_a \Big)$ are three different ordering wavevectors $\boldsymbol{Q}_a$, $\boldsymbol{Q'}_a$, $\boldsymbol{Q''}_a$ and relative phase $\gamma_a$ at a sublattice $a$; For sublattice $a=0,1,2$ and $3$, 
$\Big( \boldsymbol{Q_2}, \boldsymbol{Q_3}, \boldsymbol{Q_4}, 0 \Big)$, 
$\Big( \boldsymbol{Q_1}, \boldsymbol{Q_4}, \boldsymbol{Q_3}, \pi \Big)$, 
$\Big( \boldsymbol{Q_4}, \boldsymbol{Q_1}, \boldsymbol{Q_2}, \pi \Big)$ and
$\Big( \boldsymbol{Q_3}, \boldsymbol{Q_2}, \boldsymbol{Q_1}, \pi \Big)$ respectively where ordering wavevectors $\boldsymbol{Q}_s$ are symmetry related distinct $L$ points defined by 
$\boldsymbol{Q}_1\!=\!(\pi,\pi,\pi)$, $\boldsymbol{Q}_2\!=\!(-\pi,\pi,\pi)$,
$\boldsymbol{Q}_3\!=\!(\pi,-\pi,\pi)$ and $\boldsymbol{Q}_4\!=\!(\pi,\pi,-\pi)$.
As described in Eq.\eqref{eq:cubic}, spin configuration at every sublattice is described by three distinct ordering wavevectors and each ordering wavevector describes $x$,$y$ or $z$ components of spin for particular sublattice. One can easily understand the magnetic unit cell size is octupled compared to the original FCC lattice unit cell. With an infinitesimal $J_1$, the {\it `cubic'} phase undergoes the {\it `distorted cubic'} phase, discussed in the following paragraph.
%
%

The parameter change of $\phi$ from 0 makes small deviation of spin direction from the vertices of cube, resulting in {\it `distorted cubic'} phase, shown in parameter space colored with purple in the phase diagram Fig.\ref{fig:phase-diagram}. In this phase, spins at each sublattice point 8 different directions forming a rhombohedron. For instance, the cubic structure formed by the common origin plot of spins at sublattice $0$ (colored in blue in Fig.\ref{fig:distorted-cubic}) goes through the rhombohedral distortion along $[111]$ direction. Similarly, the cubic structure formed by the spins at different sublattices also goes through the rhombohedral distortion along their local ${\hat z}$ axis. 
Similar to the {\it cubic} phase, magnetic ordering wavevectors are at high symmetric $L$ points with octupled magnetic unit cell. 
This {\it `distorted cubic'} phase is remarkably stable against the AF NN exchanges. It can be understood by the fact that the net sum of spins for each tetrahedron is quite small in this phase, thus it is stable even in the presence of large $J_1\!>\!0$.

In these phases, one can also expect the finite scalar spin chirality as we discussed in the case of the {\it `octagonal prism'} state and it could generate the  orbital current and orbital magnetization even in Mott insulators. However, unlike the case of {\it `octagonal prism'} state, we found that the induced orbital current in these phases does not open special channels for uniform current flow along certain directions. \\ 

 {{\it Application to spinel compound GeCo$_2$O$_4$ }} --- In spinel compounds GeCo$_2$O$_4$, A-site Ge$^{4+}$ is non-magnetic and B-site Co$^{2+}$ ions have partially filled $d$ orbitals with $3d^7$ surrounded by octahedrally coordinated oxygens. It is known that Co$^{2+}$ realizes high spin states with $S\!=\!3/2$ due to large Hund's coupling. Thus, five electrons are occupied in $t_{2g}$ orbitals and two electrons are occupied in $e_g$ orbitals. 
Within $t_{2g}$ orbitals, five electrons in the presence of spin orbit coupling give rise to $j_{\text{eff}}\!=\!1/2$ configuration and thus spin dependent hopping of electrons with $j_{\text{eff}}\!=\!1/2$ is quite important via superexchange mechanism in spinels.
 
 In GeCo$_2$O$_4$, the magnetic ordering occurs at $T_{\text N} \! = \! 21$K and the neutron powder diffraction experiments exhibit the Bragg peak at $\pi(1,1,1)$ due to magnetic reflection.\cite{diaz2006magnetic,matsuda2011magnetic,fabreges2017field} In addition to this Bragg peak, there exist additional peaks at $\pi(\!-\!1,1,1)$, $\pi(1,\!-\!1,1)$ and $\pi(1,1,\!-\!1).$\cite{matsuda2011magnetic} These wavevectors are nothing but ${\boldsymbol{Q}_1}$, ${\boldsymbol{Q}_2}$, ${\boldsymbol{Q}_3}$ and $\boldsymbol{Q}_4$ defined to describe the {\it `distorted cubic'} phase.
 In this material, there is another issue related to concurrent structural transition, which could be influential to determine magnetic ordering. However, more recent experimental studies have been resolving this issue by observing two separate transitions; structural transition at $T_{\text S} \! =\!16$K and magnetic ordering temperature at $T_{\text N} \! =\! 21$K.\cite{barton2014structural} 
Furthermore, the magnetic ordering is known to be non collinear based on magnetoelectric signature.\cite{seshadri2017} Thus, one can consider such magnetic ordering is not originated from the structural transition and we argue that the {\it `distorted cubic'} phase could be relevant to explain the magnetic ordering observed in GeCo$_2$O$_4$. For more argument, however, future experiments will be required to analyze the exact magnetic ground state. \\

{\bf Discussion} 

In this paper, we have studied the generic spin model of pyrochlore spinels. Motivated by geometrical frustration in pyrochlores, we further explored the role of anisotropic spin interactions. Especially focusing on the spin orbit coupled system, we derived the spin exchanges based on psudospin $j_{\rm{eff}} \!=\!1/2$ Hubbard model. It turns out that the anisotropic spin exchanges between next-nearest-neighbors play a dominant role in determining magnetic ground state due to complicated superexchange paths in pyrochlore spinels. It is notable that {\it local XY}, {\it all-in-all-out} and {\it Palmer-Chalker} phases described by zero ordering wavevectors also emerge purely by anisotropic exchanges between next nearest neighbor even in the absence of nearest neighbor interactions. More remarkably, we have newly discovered {\it non-coplanar magnetic orderings} described by finite ordering wave vectors; {\it `octagonal prism'} phase and {\it `(distorted) cubic'} phase. These non-coplanar states can give rise to unique properties in the Mott insulators, such as orbital current and orbital magnetization induced by scalar spin chirality. One of the pronounced effect in the {\it `octagonal prism'} state is an alternating net current along certain direction. In addition, we also discuss the {\it `distorted cubic'} phase may help understanding the magnetic order in GeCo$_2$O$_4$ spinel compounds. Our theoretical prediction of exotic magnetic states opens new search of pyrochlore spinel materials with spin orbit coupling, and guide further theoretical and experimental studies.    \\

\vskip 0.2cm 

{\bf Acknowledgments }

We would like to thank L. Balents, H. Takagi, R. Seshadri and S. Ji for useful discussions and comments. 
Hospitality at APCTP during the program ``Asia Pacific Workshop on Quantum Magnetism" is kindly acknowledged. S.B.L thanks the hospitality at the Physics Department of University of California, San Diego. 
G.B.S and S.B.L are supported by the KAIST startup funding and National Research Foundation Grant (NRF-2017R1A2B4008097). \\

\newpage
\newpage

\section{Slater-Koster parametrization}


In particular, pseudospin $j_{\text{\it eff}}$=$1/2$ is characterized by the mixture of spin $\uparrow$, $\downarrow$ and $t_{2g}$ orbitals $ | j^z_{\text{\it eff}}$ = $1/2 \ra $ =  $ \Big( | yz,\! \downarrow \ra \! + i | xz, \downarrow \ra \! +  |xy,\! \uparrow \!  \ra \! \Big) \!/ \! \sqrt{3}$, $ | j^z_{\text{\it eff}}$ = $- 1/2 \ra $ =  $ \Big( | yz,\! \uparrow \ra \! - i | xz, \uparrow \ra \! -   |xy,\! \downarrow \!  \ra \! \Big) \!/ \! \sqrt{3}$. Then spin projection can be represented within below equation.

\bea
\left( \!\!\! 
\begin{array}{c} | j^z_{\text{\it eff}} \!=\!  1\!/2 \ra  \\ | j^z_{\text{\it eff}} \!=\!-\!1\!/2 \ra  \\ \end{array} \!\!\! \right)
\!=\!
\left(\begin{array}{cccccc} \!0\! & \! \frac{1}{\sqrt{3}}\!  & \! 0\!  & \! \frac{i}{\sqrt{3}} \! & \! \frac{1}{\sqrt{3}}\!  & \! 0 \! \\ 
	\! \frac{1}{\sqrt{3}}\!  & \! 0\!  & \! \frac{\! -\! i}{\sqrt{3}}\!  & \! 0\!  &\!  0\!  & \!  \frac{\! -\! 1}{\sqrt{3}}\!  \\\end{array}\right) \!\!
\left(\! \! \begin{array}{c} | yz,\! \uparrow \ra  \\ | yz,\! \downarrow \ra  \\  | xz,\! \uparrow \ra  \\ | xz,\! \downarrow \ra  \\ | xy,\! \uparrow \ra \\ | xy,\! \downarrow \ra \\\end{array} \! \! \right).
\label{eq:projection}
\eea

Let us exemplify how hopping magnitudes vanish between NN. We focus on two sublattices, B$_0$ at $(x_0,y_0,z_0)$ and sublattice B$_3$ at $(x_3,y_3,z_3)$ where $z_0\!\!=\!\!z_3$. Then, there exist two superexchange paths through X$_1$ and X$_2$ sites located at $(x_3,y_0,z_0)$ and $(x_0,y_3,z_0)$ respectively. In the former case, $d_{xz}$(B$_0$)-$p_z$(X$_1$)-$d_{yz}$(B$_3$) hopping is allowed, whereas $d_{yz}$(B$_0$)-$p_z$(X$_2$)-$d_{xz}$(B$_3$) in the latter case. Within $j_{\text{\it eff}}\!\!=\!\!1/2$ doublet, each hopping has $i$ and $- i$ respectively thus they cancel with each other, resulting in vanishing hopping magnitudes between NNs. In details, we present hopping mediated via two X sites in $t_{2g}$ basis as below.

\bea
\left(
\begin{array}{cccccc}
	\! 0\!  & \! 0\!  & \! -\! (V^B_{pd\pi})^2\!  & \! 0\!  &\!  0\!  & \! 0\!  \\
	\! 0\!  & \! 0\!  &\!  0\!  & \! -\! (V^B_{pd\pi})^2\!  & \! 0\!  & \! 0\!  \\
	\! -\! (V^B_{pd\pi})^2\!  & \! 0\!  & \! 0\!  & \! 0\!  & \! 0\!  & \! 0\!  \\\
	\! 0 \! & \! -\! (V^B_{pd\pi})^2\!  & \! 0\!  & \! 0\!  & \! 0\!  &\!  0\!  \\
	\! 0\!  & \! 0\!  & \! 0\!  & \! 0\!  & \! 0\!  & \! 0\!  \\
	\! 0\!  & \! 0\!  & \! 0\!  & \! 0\!  & \! 0\!  & \! 0\!  \\\
\end{array}
\right)
\label{eq:NN-wop}
\eea

where $V^B_{pd\sigma},V^B_{pd\pi}$ represent the usual Slater-Koster parameters for direct overlap of $p$-orbital at X site and $d$-orbital at B site.
After projecting it onto  $j_{\text{\it eff}}$=$1/2$ subspace, it is written as

\bea
h_{\text{NN}}=\left(
\begin{array}{cc}
	0 & 0 \\
	0 & 0 \\
\end{array}
\right).
\label{eq:NN}
\eea

Since above procedure can be applied to any NN sites with same local environment, it definitely shows that B$_a$-X-B$_b$ hopping magnitude vanishes for any B$_{a}$ and B$_{b}$.

%
%

In contrast to the case of NNs, there exist non-vanishing superexchange paths for the case of NNNs. For clarification, let us consider site $i$ at $(x_{0},y_{0},z_{0})$ and $j$ at $(x_{0}-\frac{1}{4},y_{0}+\frac{1}{4},z_{0}+\frac{1}{2})$ as above. Then we look into two path B$_0$-X$_n$-A$_m$-X$_l$-B$_3$ where $\big\{\text{X}_n,\text{A}_m,\text{X}_l\big\}$ are located at $\big\{(x_{0},y_{0},z_{0}+1/4),(x_{0}-1/8,y_{0}-1/8,z_{0}+3/8),(x_{0}-1/4,y_{0},z_{0}+1/2)\big\}$ and $\big\{(x_{0}-1/4,y_{0},z_{0}),(x_{0}-3/8,y_{0}+1/8,z_{0}+1/8),(x_{0}-1/4,y_{0}+1/4,z_{0}+1/4)\big\}$ for each path. Considering that A site is occupied by ions filled with 3d orbitals, we get a hopping matrix, mediated by $t_{2g}$ levels at A sites, in a projected $j_{\text{\it eff}}$=$1/2$ basis as following.

\bea
h_{\text{NNN}}^{\text{A}_{t_{2g}}}=\left(
\begin{array}{cc}
	V^{t_{2g}}_{11} & V^{t_{2g}}_{12} \\
	V^{t_{2g}}_{21} & V^{t_{2g}}_{22} \\
\end{array}
\right).
\label{eq:NNN-t2g}
\eea

Each components of the matrix are given by
\begin{widetext}
	\bea
	V^{t_{2g}}_{11} &=&\frac{2+2i}{81} (V^B_{pd\pi })^2 \left( \sqrt{3}(3+i) V^A_{pd\pi } V^A_{pd\sigma}+ (3+2i) (V^A_{pd\pi })^2 -3i (V^A_{pd\sigma })^2\right),
	\\
	V^{t_{2g}}_{12}&=&\frac{1-i}{81} (V^B_{pd\pi })^2 \left(-2\sqrt{3} V^A_{pd\pi } V^A_{pd\sigma}+ 5(V^A_{pd\pi })^2+ 6 (V^A_{pd\sigma })^2\right),
	\\
	V^{t_{2g}}_{21}&=&\frac{1+i}{81} (V^B_{pd\pi })^2 \left( 2\sqrt{3} V^A_{pd\pi } V^A_{pd\sigma}-5 (V^A_{pd\pi })^2 -6 (V^A_{pd\sigma })^2\right),
	\\
	V^{t_{2g}}_{22}&=&\frac{-2-2i}{81} (V^B_{pd\pi })^2 \left( \sqrt{3}(1+3i) V^A_{pd\pi } V^A_{pd\sigma}+(2+3i) (V^A_{pd\pi })^2 -3(V^A_{pd\sigma })^2\right)
	\eea
\end{widetext}

where $V^A_{pd\sigma},V^A_{pd\pi}$ represent direct overlap of $p$-orbital at X site and $d$-orbital at A site.
Similarly, when we consider hoppings mediated by $e_{g}$ levels at A sites, we get a matrix as
\bea
h_{\text{NNN}}^{\text{A}_{e_{g}}} \!=
\! \left(\!\!\!
\begin{array}{cc}
	\frac{4\!-\!4i}{27} (V^B_{pd\pi }V^A_{pd\pi })^2 & \frac{-2\!+\!2i}{27} (V^B_{pd\pi }V^A_{pd\pi })^2 \\
	\frac{2\!+\!2i}{27} (V^B_{pd\pi }V^A_{pd\pi })^2 & \frac{4\!+\!4i}{27} (V^B_{pd\pi }V^A_{pd\pi })^2 \\
\end{array}
\!\!\! \right).
\label{eq:NNN-eg}
\eea

\begin{figure} [!h]
	\includegraphics[width=0.8\columnwidth]{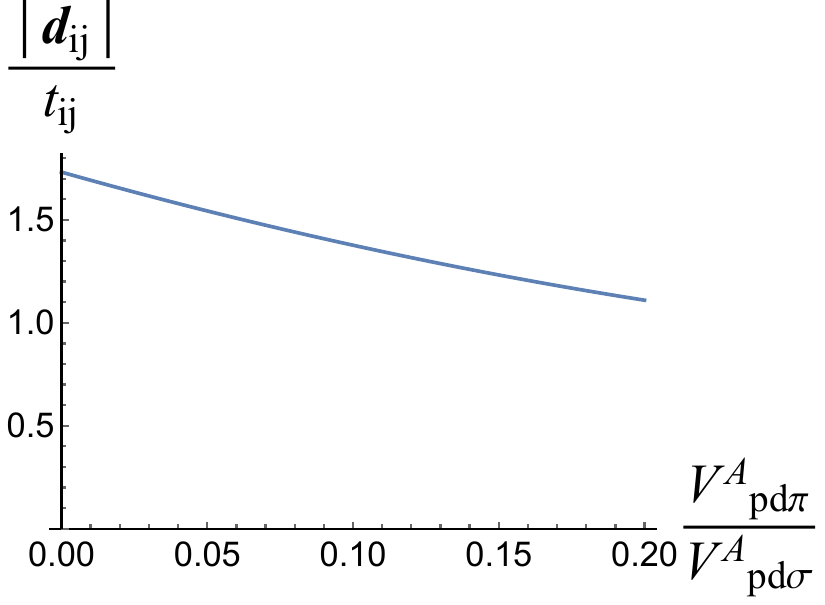}
	\caption{(color online) Ratio between spin independent and dependent hoppings for NNN via A-site $t_{2g}$ orbitals.}
	\label{fig:oaot2g}
\end{figure}

\begin{figure} [!h]
	\includegraphics[width=0.8\columnwidth]{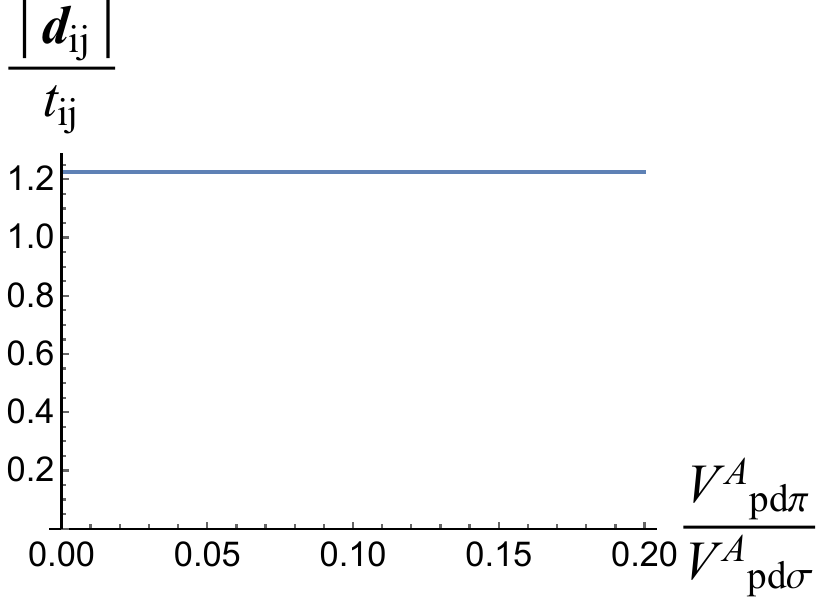}
	\caption{(color online) Ratio between spin independent and dependent hoppings for NNN via A-site $e_{g}$ orbitals.}
	\label{fig:oaoeg}
\end{figure}

Since $h_{NNN}^{\text{A}t_{2g}}$ and $h_{NNN}^{\text{A}e_{g}}$ can be directly mapped onto $t_{ij}$ and $\textbf{d}_{ij}$ in main text, we plot $|\textbf{d}_{ij}|/t_{ij}$ as a function of $V^A_{pd\pi }/V^A_{pd\sigma }$ for both cases. Fig.\ref{fig:oaot2g} and Fig.\ref{fig:oaoeg} clearly demonstrate the importance of spin dependent hopping between NNNs.

\section{Spatial symmetry constraint on Model}

We start from simple Hubbard model with hopping between site $i$ and its next-nearest-neighbor (NNN) site $j$.

\bea
\mathcal H = \sum_{\la \la ij \ra \ra} {\mathcal H}_{ij}+U \sum_i n_{i\uparrow}n_{j\downarrow},
\label{eq:Hubbard}
\eea

where ${\mathcal H}_{ij} \equiv c^\dagger_{i \alpha} \Big( t_{ij} \mathbb{I}+ i {\textbf{d}}_{ij} \cdot \boldsymbol{\sigma}  \Big)_{\alpha \beta} c_{j \beta} $.

First we notice that NNN path between $i$ site and $j$ site should pass through $k$ site which is common NN of previous two sites. Then $\mathcal H_{ij}$ has a $C_2$ symmetry along normal vector $\hat{\textbf{n}}_{ij} \equiv \frac{\textbf{r}_{ik}-\textbf{r}_{kj}}{\left\| \textbf{r}_{ik}-\textbf{r}_{kj}\right\|}$, where $\textbf{r}_{ij} \equiv \textbf{r}_i-\textbf{r}_j$. Under rotation operator $\textbf{R}(\hat{\textbf{n}}_{ij},\pi)$, which rotate angle $\pi$ about $\hat{\textbf{n}}_{ij}$ axis, creation and annihilation operator transform as

\bea
c^\dagger_{i\alpha} &\to& \sum_{\beta} c^\dagger_{j\beta} (\exp(\frac{-i\pi\hat{\textbf{n}}_{ij}\cdot\boldsymbol{\sigma}}{2}))_{\beta\alpha}, \nonumber \\~
c_{i\alpha} &\to& \sum_{\beta} \exp(\frac{i\pi\hat{\textbf{n}}_{ij}\cdot\boldsymbol{\sigma}}{2})_{\alpha\beta} c_{j\beta}.
\label{eq:unitary}
\eea

Hence $\mathcal H_{ij}$ transform as following. 

\bea
c^\dagger_{i \alpha} \Big( t_{ij} \mathbb{I} \!+\! i {\textbf{d}}_{ij} \cdot \boldsymbol{\sigma}  \Big)_{\alpha \beta} c_{j \beta} &
\nonumber \\ ~\to~  c^\dagger_{i \alpha} (-i  \hat{\textbf{n}}_{ij} \cdot \boldsymbol{\sigma}) \Big( t_{ij} \mathbb{I} & \!+\! i {\textbf{d}}_{ij} \cdot \boldsymbol{\sigma}  \Big)_{\alpha \beta}  (i  \hat{\textbf{n}}_{ij} \cdot \boldsymbol{\sigma}) c_{j \beta}.
\label{eq:unitary-Hamiltonian}
\eea

Meanwhile, $\mathcal H_{ij}$ should be invariant under $\textbf{R}(\hat{\textbf{n}}_{ij},\pi)$ operator which leads to

\bea
\hat{\textbf{n}}_{ij} \cdot  \hat{\textbf{d}}_{ij}=0.
\label{eq:d-sym}
\eea

Similarly, for the case of nearest-neighbor (NN), associated two mirror symmetry operators constrain $\hat{\textbf{d}}_{ij}$ to be written as

\bea
\hat{\textbf{d}}_{ij}=\frac{\textbf{r}_{kl}}{\left\| \textbf{r}_{kl}\right\|}
\label{eq:d-NN}
\eea  

where $k$ and $l$ correspond to common NN of $i$ site and $j$ site. 

With $U$ much greater than hopping amplitude, $t$ and $\textbf{d}$, we can derive effective spin Hamiltonian using 2nd order perturbation theory for half-filled case. Treating hopping terms perturbatively, we get effective low energy spin Hamiltonian
\bea
{\mathcal H}_{ij} \to &\frac{t_{ij}^2+d_{ij}^2}{U} \Big((\cos{2\theta})(\textbf{S}_i\cdot\textbf{S}_j)\mp(\sin{2\theta})\hat{\textbf{d}}_{ij}\cdot(\textbf{S}_i\times\textbf{S}_j) \nonumber \\
& +(1-\cos{2\theta})(\textbf{S}_i\cdot\hat{\textbf{d}}_{ij})(\textbf{S}_j\cdot\hat{\textbf{d}}_{ij}) \Big),
\label{eq:spin-model}
\eea

where $\theta = \tan^{-1} \Big( \pm | \textbf{d}_{ij} | /t_{ij} \Big) $.
Above perturbative treatment can be applied for the case of NNs in which spin Hamiltonian has free parameter $\Omega$ instead of $\theta$.

To write $\hat{\textbf{d}}_{ij}$ for a given bonding, we focus on one of NNN bondings connecting site $i$ at $(x_0,y_0,z_0)$, which belong to sublattice B$_0$, and site $j$ at $(x_{0}-\frac{1}{4},y_{0}+\frac{1}{4},z_{0}+\frac{1}{2})$, which belong to sublattice B$_3$. This bond has its normal vector $\hat{\textbf{n}}_{ij}=(-\frac{1}{\sqrt{2}},-\frac{1}{\sqrt{2}},0)$. Then we are able to parametrize $\hat{\textbf{d}}_{ij}$ as
\bea
\hat{\textbf{d}}_{ij}=(-\frac{\sqrt2}{2}\sin{\phi},\frac{\sqrt2}{2}\sin{\phi},-\cos{\phi}).
\label{eq:dij}
\eea
We note that $\hat{\textbf{d}}_{ij}$ for other NNN bonds can be also determined by this parameter $\phi$ on symmetry grounds.
By plugging in Eq.\ref{eq:dij} to Eq.\ref{eq:spin-model} with $-$ sign, we get

\bea
\textbf{J}_{03}(\phi,\theta) = 
\left(
\begin{array}{ccc}
	j_6 & j_3 & -j_2 \\
	j_5 & j_6 & -j_4 \\
	j_4 & j_2 & j_1 \\
\end{array}
\right)
\label{eq:J03}
\eea

where subscript of $\textbf{J}$, $03$, indicate the bonding between B$_0$ and B$_3$ (pyrochlore sublattices of B-site spinel).

In Eq.\ref{eq:J03}, $j_n$ are given by 

\bea
j_1&=& \frac{t_{ij}^2\!+\!d_{ij}^2}{U} \Big((1\!-\!\cos{2 \theta} ) \cos^2{\phi} +\cos{2 \theta} \Big), \nonumber \\
j_2 &=& \frac{t_{ij}^2\!+\!d_{ij}^2}{U} \Big(\!-\!\frac{\sin{2 \theta} \sin{\phi}}{\sqrt{2}} \!-\!\frac{(1\!-\! \cos{2 \theta}) \sin{\phi} \cos{\phi}}{\sqrt{2}}\Big), \nonumber \\
j_3 &=& \frac{t_{ij}^2\!+\!d_{ij}^2}{U}(\sin{2 \theta} \cos{\phi} \!-\!\frac{1}{2} (1\!-\!\cos{2 \theta}) \sin^2{\phi}), \nonumber \\
j_4 &=& \frac{t_{ij}^2\!+\!d_{ij}^2}{U} \Big(\frac{(1\!-\! \cos{2 \theta}) \sin{\phi} \cos{\phi}}{\sqrt{2}}-\frac{\sin{2 \theta} \sin{\phi}}{\sqrt{2}} \Big), \nonumber\\
j_5 &=& \frac{t_{ij}^2\!+\!d_{ij}^2}{U}\Big(\!-\!\frac{1}{2} (1\!-\! \cos{2 \theta}) \sin ^2{\phi}\!-\!\sin{2 \theta} \cos{\phi} \Big), \nonumber\\
j_6 &=& \frac{t_{ij}^2\!+\!d_{ij}^2}{U} \Big(\frac{1}{2} (1\!-\! \cos{2 \theta}) \sin ^2{\phi}+\cos{2 \theta} \Big).
\label{eq:j03}
\eea

Using combinations of spatial symmetries, one can parametrize $\textbf{J}_{ab}(\phi,\theta)$ for any NNN bondings where $a$ and $b$ indicate sublattices to which sites $m$ and $n$ each belongs. One thing to notice is that there exists two kinds of $\textbf{J}_{ab}$ for every sublattice pair $ab$. (They are related by mirror symmetries.) Let's consider a site $i$ at $(x_0,y_0,z_0)$, which belong to sublattice B$_0$, and a site $k$ at $(x_{0}+\frac{1}{4},y_{0}-\frac{1}{4},z_{0}+\frac{1}{2})$, which belong to sublattice B$_3$. Then $\textbf{J}_{03}$ for sites $i$ and $k$ is written as

\bea
\textbf{J}_{03}(\phi,\theta) = 
\left(
\begin{array}{ccc}
	j_6 & j_5 & -j_4 \\
	j_3 & j_6 & -j_2 \\
	j_2 & j_4 & j_1 \\
\end{array}
\right).
\label{eq:J03'}
\eea

Same procedure for NN can be applied starting from Eq.\ref{eq:spin-model} with $-$ sign. If we consider two sublattices, B$_0$ and B$_3$, in a unit cell as shown in the main text, we get

\bea
\textbf{J}_{03}^{'}(\phi,\theta) = 
\left(
\begin{array}{ccc}
	j_2^{'} & j_3^{'} & j_4^{'} \\
	j_3^{'} & j_2^{'} & j_4^{'} \\
	-j_4{'} & -j_4^{'} & j_1^{'} \\
\end{array}
\right),
\label{eq:J'03}
\eea

and its components are given by

\bea
j_1^{'} &=& \cos (2 \Omega ),\nonumber\\
j_2^{'}&=& \frac{1}{2} \cos (2 \Omega )+\frac{1}{2},\nonumber\\
j_3^{'}&=& \frac{1}{2} \cos (2 \Omega )-\frac{1}{2},\nonumber\\
j_4^{'}&=& -\frac{\sin (2 \Omega )}{\sqrt{2}}.
\eea

Unlike the case of NNNs, there exists only one kind of $\textbf{J}_{ab}^{'}$ for every sublattice pair $ab$.

We can also see how spatial symmetries of each bond are taken into account starting from the most general form of NNN spin Hamiltonian,

\bea
{\mathcal H} = \sum_{\la \la ij \ra \ra} J_{ij}^{\alpha\beta}(S_i^\alpha S_j^\beta).
\label{eq:spin-Hamiltonian}
\eea

As before we look into one of NNN bondings connecting sites $i$ at $(0,0,0)$ and $j$ at $(-\frac{1}{4},\frac{1}{4},\frac{1}{2})$. Under $\textbf{R}(\hat{\textbf{n}}_{ij},\pi)$ operator each spin component transform as

\bea
\left (\begin{array}{ccc} S_i^x \\ S_i^y \\  S_i^z \end{array} \right) \to \left(\begin{array}{ccc} S_j^y \\S_j^x \\ - S_j^z \end{array} \right),~ \left (\begin{array}{ccc} S_j^x \\ S_j^y \\ S_j^z \end{array} \right) \to \left(\begin{array}{ccc}  S_i^y \\  S_i^x \\ -S_i^z \end{array} \right).
\label{eq:spin-sym}
\eea

Since ${\mathcal H}_{ij}$ should be invariant under $\textbf{R}(\hat{\textbf{n}}_{ij},\pi)$ operator and three equalities emerge as

\bea
J_{ij}^{zy} = -J_{ij}^{xz},~J_{ij}^{zx} = -J_{ij}^{yz},~J_{ij}^{yy}= J_{ij}^{xx}.
\label{eq:Jij-sym}
\eea

By taking account of above equalities, we parametrize $\textbf{J}_{ij}$ as

\bea
\textbf{J}_{ij} = 
\left(
\begin{array}{ccc}
	j_6 & j_3 & -j_2 \\
	j_5 & j_6 & -j_4 \\
	j_4 & j_2 & j_1 \\
\end{array}
\right)
\label{eq:Jij}
\eea

Simultaneously, any $\textbf{J}_{kl}$ can be parametrized in terms of $j_i$ with $i$ 1 to 6 using spatial symmetry operator as before.
Similar argument has been done for NN case which states that one need 4 independent parameters to describe spin Hamiltonian.

\section{Luttinger-Tisza}

There is a systematic method, which helps us to find classical magnetic ground state, called Luttinger-Tisza (LT) method. In a specific spin model, LT method find ordering wavector with so called "weak constraint", $\sum_i|\textbf{S}_i|^2=NS$, where $N$ is number of sites.
We start from a general spin Hamiltonian,
\bea
\mathcal H = \sum_{ij} J_{ij}^{\alpha\beta} S_i^\alpha S_j^\beta
\label{eq:LT-spin}
\eea

where $i$ and $j$ are site indicies while $\alpha$ and $\beta$ run over three components of spin. If we do the Fourier transformation, we do get

\bea
\tilde{\textbf{S}}_a (\textbf q) = \sum_{i\in a}\textbf{S}_i e^{i\textbf{q}\cdot\textbf{r}_i}, ~\tilde{J}_{ab}^{\alpha\beta} (\textbf q) = \sum_{i\in a}\sum_{j\in b}J_{ij}^{\alpha\beta} e^{i\textbf{q}\cdot\textbf{r}_{ij}}
\label{eq:LT-fourier}
\eea

where $a$ and $b$ indicate sublattice indicies.

Then the Hamiltonian becomes

\bea
\mathcal H =\sum_{ab}\sum_{\alpha\beta}\sum_{\textbf q}\tilde{J}_{ab}^{\alpha\beta} (\textbf q) \tilde{S}_a^\alpha (-\textbf q)\tilde{S}_b^\beta (\textbf q).
\label{eq:LT-Hfourier} 
\eea

\begin{figure} [!h]
	\includegraphics[width=0.8\columnwidth]{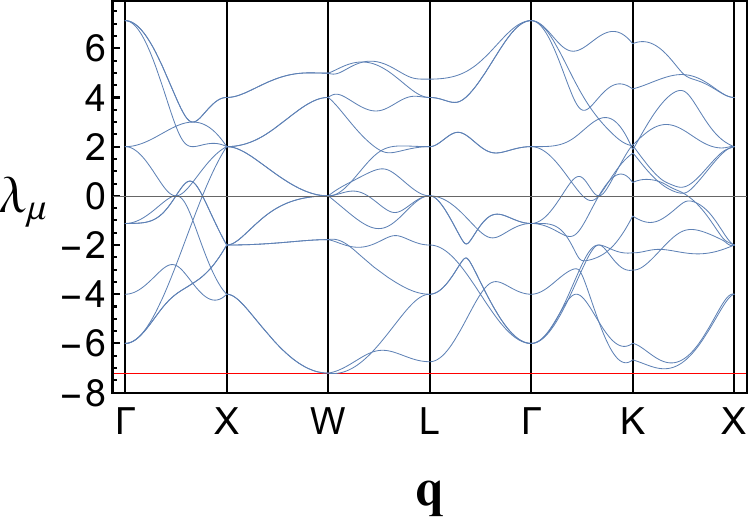}
	\caption{(color online) LT spectrum for {\it `octagonal prism'} : With ($\phi\!=\!\pi -\tan^{-1}\! \left(\sqrt{2}\right)$, $\theta \! =\! \frac{2\pi}{3}$), $\textbf Q_{LT}$ lies at high symmetric W point.}
	\label{fig:LTW}
\end{figure}
\begin{figure} [!h]
	\includegraphics[width=0.8\columnwidth]{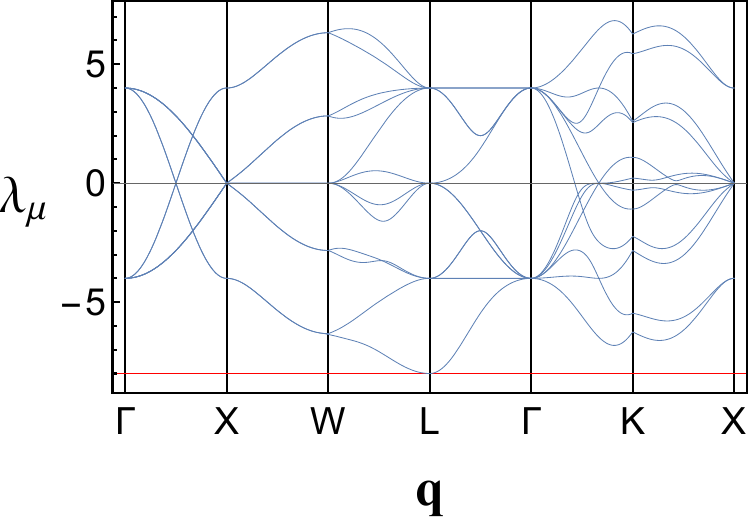}
	\caption{(color online) LT spectrum for {\it `(distorted) cubic'} : With ($\phi \!=\! 0,\theta \!=\! \frac{3\pi}{4}$), $\textbf Q_{LT}$ lies at high symmetric L point.}
	\label{fig:LTL}
\end{figure}

On a pyrochlore lattice, we get $12\times12$ matrix, $\tilde{\textbf{J}}(\textbf q)$, which leads to 12 eigenmodes corresponding to individual eigenvalues, $\lambda_\mu(\textbf{q})$. We label minimum eigenvalue among 12 as $\lambda_{LT}(\textbf{q})$. And we refer 12 components eigenmodes, whose eigenvalue is $\lambda_{LT}(\textbf{q})$, as LT eigenmodes. Next we look for optimal LT wavevector, $\textbf Q_{LT}$, which has lowest eigenvalue, $\lambda_{LT}(\textbf Q_{LT})$, among $\lambda_{LT}(\textbf{q})$ by sweeping $\textbf{q}$ in first Brillouin zone. For bipartite lattice with Heisenberg interactions, we can easily construct spin configuration, which is a linear superposition of LT eigenmodes, that satisfies strong constraint€, $|\textbf{S}_i|^2=1$. But for a pyrochlore lattice it is not trivial since LT eigenmodes do not always satisfy strong constraint€ even for Heisenberg model.
For {\it `octagonal (prism)')} state and {\it `(distorted) cubic')} state, one can find two non zero $\textbf Q_{LT}$, as shown in Fig.\ref{fig:LTW} and Fig.\ref{fig:LTL} and both phases exactly satisfy strong constraint€.

\section{Iterative minimization}
As mentioned above, it is hard to find classical magnetic ground states directly from LT method that satisfy strong constraint. Thus we also adopt a simulation called iterative minimization€ which starts from random spin configuration and let it flow as a function of interaction parameters $J$. Using a general spin Hamiltonian like before, we write how each component of spin evolve in every step as,
\bea
S_i^\alpha \to S_i^\alpha - c \sum_{\beta}\sum_{j} J_{ij}^{\alpha\beta}S_j^\beta
\label{eq:IM-flow}
\eea
where $c$ is the parameter that we give in hand. We work on a pyrochlore lattice with its system size upto $12\times12\times12$ unit cells with periodic boundary condition. 
\section{Local axes}
Local axes for pyrochlore sublattices are given in Table.\ref{tab:Local}. 
        \begin{table}[h]
        \caption{ Local axes of sublattices} 
	\begin{tabular}{ |c|c|c|c|}
	        \hline
		a & ~~ $\hat{x}_a$ & ~~$\hat{y}_a$ &~~ $\hat{z}_a$    \\ \hline \hline
		0 & ~~$\frac{1}{\sqrt{2}}[\bar110]$ & ~~$\frac{1}{\sqrt{6}}[\bar1\bar12]$ & ~~$\frac{1}{\sqrt{3}}[111]$  \\ \hline
		1 & ~~ $\frac{1}{\sqrt{2}}[\bar1\bar10]$ & ~~$\frac{1}{\sqrt{6}}[\bar11\bar2]$ & ~~$\frac{1}{\sqrt{3}}[1\bar1\bar1]$  \\ \hline
		2 & ~~$\frac{1}{\sqrt{2}}[110]$ &~~ $\frac{1}{\sqrt{6}}[1\bar1\bar2]$ &~~ $\frac{1}{\sqrt{3}}[\bar11\bar1]$  \\ \hline
		3 & ~~ $\frac{1}{\sqrt{2}}[1\bar10]$ & ~~$\frac{1}{\sqrt{6}}[112]$ & ~~$\frac{1}{\sqrt{3}}[\bar1\bar11]$  \\ \hline
         \end{tabular}
	 \label{tab:Local}
	 \end{table}

%
%
%
%

\end{document}